# Unconventional superlattice ordering in intercalated transition metal dichalcogenide V$_{1/3}$NbS$_2$

Shannon S. Fender,[1] Noah Schnitzer,[2,3] Wuzhang Fang,[4] Lopa Bhatt,[5] Dingbin Huang,[6] Amani Malik,[1] Oscar Gonzalez,[1] Veronika Sunko,[6,7] Lilia S. Xie,[1,†] David A. Muller,[3,5] Joseph Orenstein,[6,7] Yuan Ping,[4] Berit H. Goodge,[8] D. Kwabena Bediako[1,9,10]*

[1] Department of Chemistry, University of California, Berkeley, CA 94720, USA
[2] Department of Materials Science, Cornell University, Ithaca, NY, 14853, USA
[3] Kavli Institute at Cornell for Nanoscale Science, Cornell University, Ithaca, NY, 14853, USA
[4] Department of Materials Science and Engineering, University of Wisconsin, Madison, WI, 53706, USA
[5] School of Applied and Engineering Physics, Cornell University, Ithaca, NY, 14853, USA
[6] Material Sciences Division, Lawrence Berkeley National Laboratory, Berkeley, CA 94720, USA
[7] Department of Physics, University of California, Berkeley, CA 94720, USA
[8] Max-Planck-Institute for Chemical Physics of Solids, Nöthnitzer Str. 40, 01187, Dresden, Germany
[9] Chemical Sciences Division, Lawrence Berkeley National Laboratory, Berkeley, CA 94720, USA
[10] Kavli Energy NanoScience Institute, Berkeley, CA 94720, USA

The interplay between symmetry and topology in magnetic materials makes it possible to engineer exotic phases and technologically useful properties. A key requirement for these pursuits is achieving control over local crystallographic and magnetic structure, usually through sample morphology (such as synthesis of bulk crystals versus thin-films) and application of magnetic or electric fields. Here we show that V$_{1/3}$NbS$_2$ can be crystallized in two ordered superlattices, distinguished by the periodicity of out-of-plane magnetic intercalants. Whereas one of these structures is metallic and displays the hallmarks of altermagnetism, the other superlattice, which has not been isolated before in this family of intercalation compounds, is a semimetallic noncollinear antiferromagnet that may enable access to topologically nontrivial properties. This observation of an unconventional superlattice structure establishes a powerful route for tailoring the tremendous array of magnetic and electronic behaviors hosted in related materials.

Though historically neglected as the basis for spintronic devices because of their fully compensated magnetization, antiferromagnets (AFMs) have emerged in recent years as some of the most promising platforms for next-generation magnetic data storage.[1–3] One route uses non-relativistic spin-splitting, which emerges in either collinear or noncollinear AFMs from the breaking of crystal symmetries that relate the local environments of spin-bearing atoms.[4,5] In the collinear case, termed "altermagnetism," the combination of non-relativistic spin-splitting and spin–orbit coupling (SOC) can lead to the emergence of electronic and optical properties typically associated with the lifted spin degeneracy of non-compensated magnetic materials, such as ferromagnets.[6–9] Consequently, altermagnets are currently receiving intense interest for realizing exotic phases—including the quantum anomalous Hall effect[10] and topological superconductivity[11]—as well as building blocks for spintronics.[6,9] One compound of recent interest as a candidate altermagnet is V$_{1/3}$NbS$_2$, a member of the intercalated TMDs family: T$_x$MCh$_2$ (T = first row transition metal; M = Nb, Ta; Ch = S, Se). In these materials, intercalant ions form superlattices within the ab plane that determine the extended symmetries of magnetic ions and strongly dictate their magnetic interactions.[12–14] However, the arrangement of these superlattices out-of-plane (along the c-axis) is rarely considered manipulatable. Reports on the magnetic structure of V$_{1/3}$NbS$_2$ have raised questions on whether the material hosts the A-type AFM order (spins coupled ferromagnetically in the ab plane and antiferromagnetically coupled along the c-axis) necessary to realize altermagnetism.[15–18] Recent studies suggest that the minority presence of another intercalant structure may be the key factor behind these conflicting results.[19]

Here, we isolate two bulk intercalant superlattices of V$_{1/3}$NbS$_2$, one of which is an unconventional trigonal structure (space group: $R\bar{3}c$) arising from a distinctive out-of-plane intercalant configuration compared to the traditional hexagonal lattice (space group: $P6_322$) observed in other $T_{1/3}MCh_2$ compounds. Magnetic, heat capacity, and transport measurements reveal the contrasting properties of these structures, highlighting the role of both in-plane and now, out-of-plane, superlattice engineering in shaping the magnetism and topology of V$_{1/3}$NbS$_2$.

Two batches of V$_{1/3}$NbS$_2$ single crystals were synthesized by chemical vapor transport. Batches I and II were prepared with molar ratios of 1 V: 3 Nb: 6 S and 1.15 V: 3 Nb: 6 S, respectively. Additional synthetic details are provided in Materials and Methods (S1). Single-crystal X-ray diffraction (SCXRD) analysis of Batch I indicates that crystals adopt the expected $P6_322$ space group, with lattice parameters $a = b = 5.7329(10)$ Å, $c = 12.098(3)$ Å and angles $\alpha = \beta = 90°$ and $\gamma = 120°$. Figure 1a depicts two illustrated views of the $P6_322$ structure, conventional for $T_{1/3}MCh_2$, with the characteristic out-of-plane AB sequence of an in-plane $\sqrt{3}a_0 \times \sqrt{3}a_0$ ($\sqrt{3} \times \sqrt{3}$) superlattice, where $a_0$ is the lattice parameter of parent NbS$_2$. In the experimental refinement, moderate disorder is detected in the interlamellar region, as discussed in Section S2 of the Supplementary Information. Atomic-resolution high-angle annular dark field scanning transmission electron microscopy (HAADF-STEM) imaging (Figure 1b and Figure S1) shows ordered regions with the expected out-of-plane AB sequence along with a distinct ABC sequence of 'stacking faults' that occasionally disrupts the periodicity of the $P6_322$ space group.

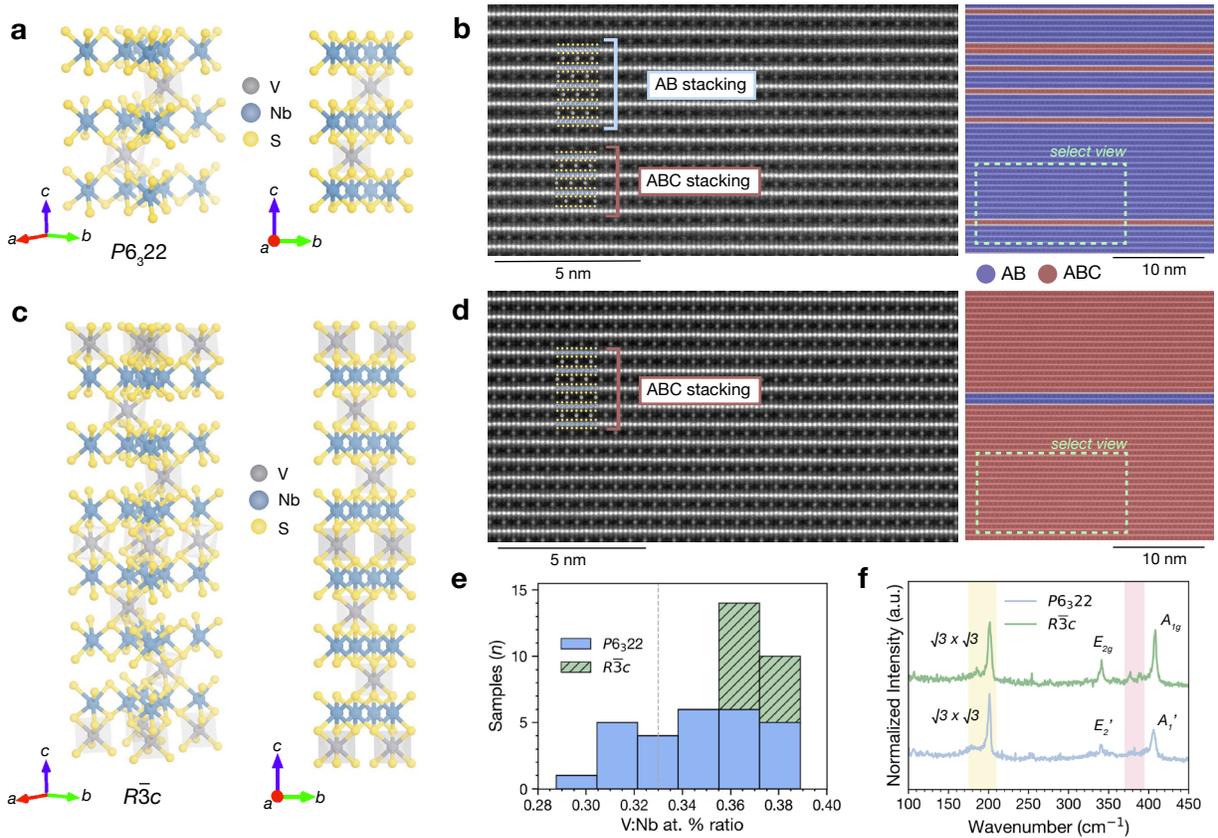

**Figure 1. a)** Two views of the crystal structure of V$_{1/3}$NbS$_2$ in a $P6_322$ space group. **b)** HAADF-STEM image of a $P6_322$ crystal projected along the $[10\bar{1}0]$ direction. **c)** Two views of crystal structure of V$_{1/3}$NbS$_2$ in a $R\bar{3}c$ space group. **d)** HAADF-STEM of a $R\bar{3}c$ crystal along $[10\bar{1}0]$. In **b** and **d**, schematic overlays in the left panels illustrate AB and ABC intercalant arrangements, while the right panels map their spatial distribution across wider fields of view. **e)** Normalized V:Nb atomic percentages (collected by EDS) of single crystals synthesized in this study. **f)** Raman spectra of representative $P6_322$ (blue) and $R\bar{3}c$ (green) crystals of V$_{1/3}$NbS$_2$. Yellow and red bands respectively highlight intercalant phonon modes and new peaks only observed in $R\bar{3}c$.

In contrast, SCXRD refinement for Batch II (Tables S3, S4) reveals a space group of $R\bar{3}c$ (Figure 1c), which is, to our knowledge, unreported for intercalated TMDs. The $R\bar{3}c$ space group corresponds to a trigonal structure within a hexagonal setting, preserving the 2H-NbS$_2$ parent structure while exhibiting a uniform yet unconventional ABC arrangement of vanadium intercalants at octahedral sites. In this case, the lattice parameters refine to $a = b = 5.7439(11)$ Å, $c = 36.349(10)$ Å, with angles $\alpha = \beta = 90°$ and $\gamma = 120°$, reflecting the retention of an in-plane $\sqrt{3} \times \sqrt{3}$ superlattice with a nearly three-fold elongation of the unit cell along the c-axis (comprising six NbS$_2$ layers) compared to that of the $P6_322$ structure (comprising two NbS$_2$ layers), a distinction that can also be inferred in powder XRD data (Figure S3).

HAADF-STEM data (Figure 1d and Figure S2) show clear and extensive ordering of the ABC configuration seen to a lesser extent in crystals from Batch I.

Energy-dispersive X-ray spectroscopy (EDS) data collected for crystals across both growths are presented in the histogram of Figure 1e. Notably, the average vanadium content of Batch II (green) slightly exceeds that of Batch I (blue), indicating that the formation of these two polytypes may be related to the initial V:Nb stoichiometries used in crystal growth.

Room temperature Raman scattering characterization of Batch I ($P6_322$) and Batch II ($R\bar{3}c$) structures is shown in Figure 1f. The set of peaks appearing in the low-frequency regime (160–210 cm$^{-1}$) comprises the characteristic excitations of $\sqrt{3} \times \sqrt{3}$ intercalant superlattice order.[20] However, the spectrum for the $R\bar{3}c$ material is distinguished by new modes that appear between 370 cm$^{-1}$ and 400 cm$^{-1}$, with sharper spectral features overall.

To compare the magnetic properties of $P6_322$ and $R\bar{3}c$ structures of V$_{1/3}$NbS$_2$, we select crystals that exhibit compositions closest to the target $x = 1/3$ stoichiometry and vibrational characteristics consistent with the spectra shown in Figure 1f. For the $P6_322$ crystal, magnetic susceptibility, $\chi_M$, as a function of temperature (Figure 2a) reveals a phase transition around 50 K, with in-plane $\chi_M$ /mol-V values reaching a maximum of 0.069 emu/mol V Oe. This phase transition is also detected in heat capacity measurements of the same crystal (Figure 2a, inset). The absence of a sharp downturn in $\chi_M$ upon ordering (the canonical signature of antiferromagnetism) indicates uncompensated moments and weak ferromagnetism, which would be expected for an altermagnetic ground state.[21–24]

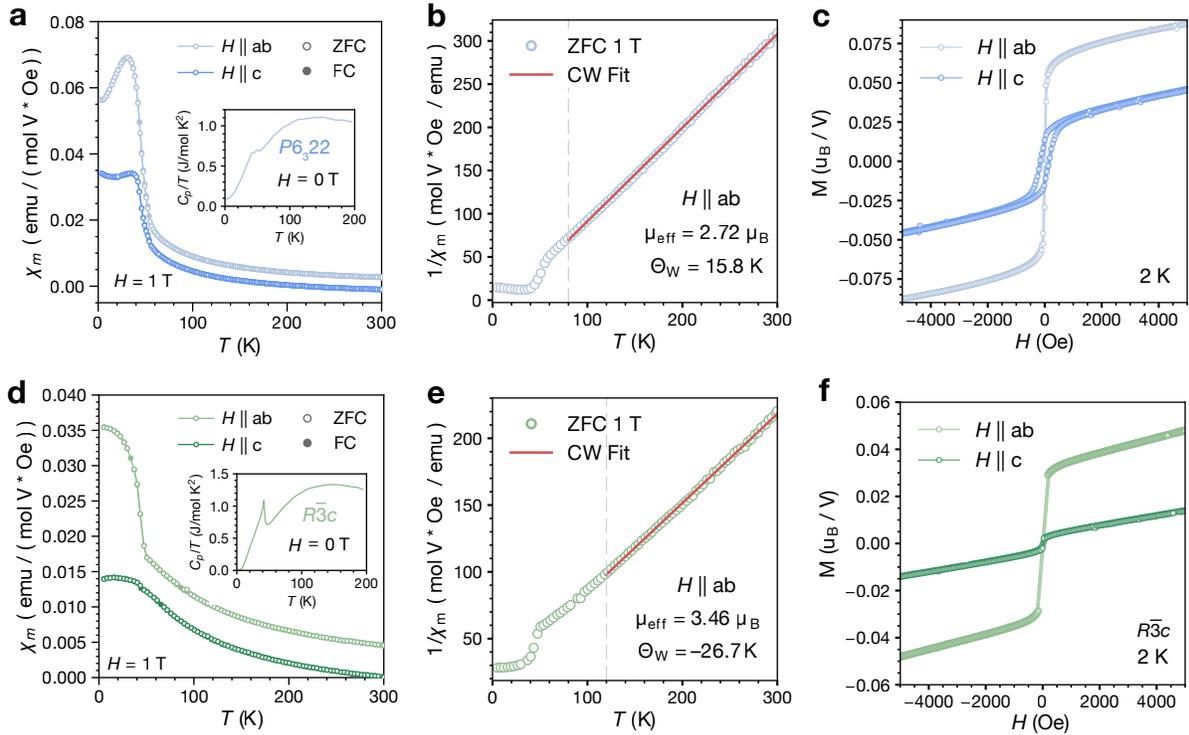

**Figure 2:** Measurements of $V_{1/3}NbS_2$ in $P6_322$ **(a–c)** and $R\bar{3}c$ **(d-f)** structures. **a,d)** Magnetic susceptibility ($\chi_M$) as a function of temperature, $T$. Inset: Heat capacity ($C_p/T$ vs. $T$) normalized to the formula $VNb_3S_6$. **b,e)** $1/\chi_M(T)$, fitted to Curie–Weiss law. **c,f)** Magnetization, $M$, as a function of field, $H$.

The out-of-plane $\chi_M$ is about twofold weaker than the in-plane $\chi_M$, consistent with magnetic moments lying in the $ab$ plane. Curie–Weiss analysis performed above 80 K and under an applied field of 1 T in the $ab$ plane yields an effective magnetic moment, $\mu_{eff}$, of 2.72 $\mu_B$/V, within 4% of the expected 2.83 $\mu_B$ for $V^{3+}$ ($d^2$) predicted by Hund's rules. The Weiss temperature, $\theta_W$, of 15.8 K is consistent with the presence of ferromagnetic interactions at high temperatures. Measurements of magnetization, $M$, as a function of field, $H$, (Figure 2c) indicate a highly compensated but weakly ferromagnetic state, which does not saturate up to 0.5 T and demonstrates minimal hysteresis.

For the $R\bar{3}c$ crystal, a clear magnetic transition onsets at 45 K, and the maximum $\chi_M$ at low temperatures (Figure 2d) is 0.035 emu/mol V Oe, half that of the $P6_322$ material. As in the $P6_322$ case, no bifurcation is observed between ZFC and FC curves. Heat capacity data acquired from the same crystal as that used for magnetism (inset of Figure 2d) show a clear phase transition also at 45 K. Curie–Weiss analysis (Figure 2e) yields $\theta_W = -26.7$ K, consistent with a predominance of antiferromagnetic interactions in the paramagnetic regime. Whereas a value of $\mu_{eff} = 3.46$ $\mu_B$/V, close to that expected for $V^{2+}$ ($d^3$)—3.87 $\mu_B$/V, is measured by magnetometry, X-ray photoemission spectra acquired from the surface of crystals, (Figure S4) appear similar for the polytypes.

For $R\bar{3}c$, measurements of $M(H)$ (Figure 2f) reveal an in-plane response comparable to that of $P6_322$. In contrast, the out-of-plane $M$ is nearly two orders of magnitude smaller and, unlike $P6_322$, exhibits no appreciable hysteresis.

Electrical transport measurements reveal the magnetoelectronic behaviors of $V_{1/3}NbS_2$ in $P6_322$ and $R\bar{3}c$ space groups. In-plane longitudinal resistivity ($\rho_{xx}$) measurements of single crystals (Figure 3a) show metallic transport responses with decreasing temperatures. The relatively sharp drops in resistivity around 45 K and 40 K for $P6_322$ and $R\bar{3}c$ crystals, respectively, are consistent with reduced carrier scattering upon magnetic ordering. Interestingly, the $R\bar{3}c$ sample shows a crossover to semiconducting behavior above 250 K.[25]

Hall resistivity ($\rho_{xy}$) values of $P6_322$ and $R\bar{3}c$ crystals (Figure 3b), are primarily governed by the ordinary Hall effect, which describes the density of free carriers.[26] Above 0.5 T, $\rho_{xy}$ data were fit using the single band expression $1/(R_H e)$; the positive $R_H$ values for both crystals indicate carrier transport dominated by holes. While the $P6_322$ crystal exhibits carrier concentrations on the order of $10^{21}$ holes/cm$^3$, consistent with other intercalated TMDs at $x=1/3$ stoichiometry,[27] the $R\bar{3}c$ crystal possesses carrier concentrations two orders of magnitude lower, ~$10^{19}$ holes/cm$^3$ (Figure 3c), which may explain the higher resistivity observed in this structure. However, we note that the resistivity in $R\bar{3}c$ crystals is only twice that of $P6_322$, implying a considerable enhancement in carrier mobility in $R\bar{3}c$ crystals (195 cm$^2$V$^{-1}$s$^{-1}$) over that in $P6_322$ (18.6 cm$^2$V$^{-1}$s$^{-1}$). The surprisingly low carrier density in $R\bar{3}c$, taken together with the crossover near 250 K from increasing to decreasing resistivity upon lowering temperature, suggests a Fermi level very close to the top of the valence band near a band gap.

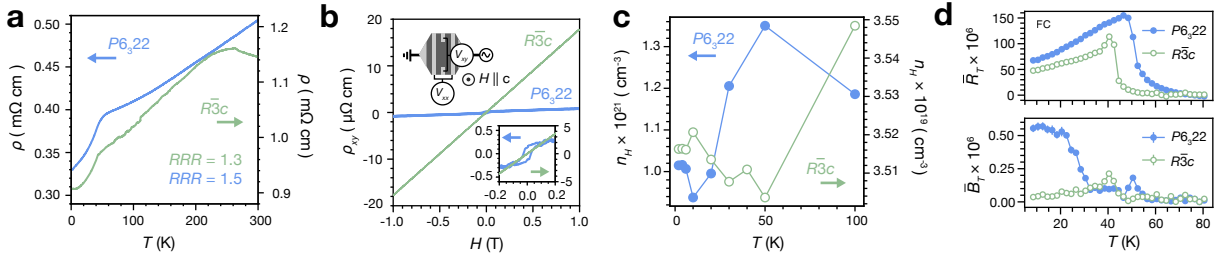

**Figure 3. a)** $T$-dependent longitudinal resistivity ($\rho_{xx}$) of $P6_322$ and $R\bar{3}c$ single crystals. **b)** Hall resistivity ($\rho_{xy}$) as a function of $H$ at 2 K. **c)** $T$ dependence of carrier concentrations ($n_H$) for $P6_322$ and $R\bar{3}c$. **d)** Normalized, thermally-modulated optical reflectivity, $\bar{R}_T$, (top) and birefringence, $\bar{B}_T$, (bottom) of $P6_322$ and $R\bar{3}c$.

These transport measurements also show that $P6_322$ crystals display a hysteretic response in $\rho_{xy}$ at low fields, the anomalous Hall effect (AHE), a key manifestation of altermagnetism.[28–30] However, no AHE is observed in $R\bar{3}c$ crystals (Figure 3b inset). As detailed in Supporting Information Section S7 (Figures S5, S6), density functional theory (DFT) calculations for $P6_322$ and $R\bar{3}c$ demonstrate altermagnetic spin-splitting along the H–K and L–F directions of their respective Brillouin zones under A-type AFM ordering. For a Néel vector along the crystallographic $a$-axis (Figure 4a), SOC induces both weak ferromagnetism and the symmetry-allowed AHE (Table S5).[31] The absence of AHE in measurements of $R\bar{3}c$ points to the adoption of a magnetic ground state distinct from that of $P6_322$.

To disambiguate the magnetic structure of the $R\bar{3}c$ polytype, Figure 3d compares the thermally modulated optical reflectivity ($\bar{R}_T$) and birefringence ($\bar{B}_T$) responses of the polytypes as a function of temperature (see SI for experimental details). Both $\bar{R}_T$ traces show peaks near $T_N$ due to changes of electronic structure at the magnetic transition in each case. In contrast, $\bar{B}_T$, which reports on rotational symmetry breaking, exhibits distinct behavior in the two polytypes. In $P6_322$, $\bar{B}_T$ rapidly increases with decreasing temperature below 35 K, in agreement with $C_{2z}$ spin-symmetries of A-type AFM order (Figure 4a,b). In contrast, in $R\bar{3}c$ there is no experimentally significant $\bar{B}_T$, allowing for spin configurations consisting of higher order $C_{nz}$ rotational symmetries ($n \geq 3$).[32] Additionally, prior neutron diffraction studies on V$_{1/3}$NbS$_2$ observed a **k** = (0,0,⅓) propagation vector that, under the assumption of a $P6_322$ crystallographic structure, was assigned to an out-of-plane canting of moments.[15] Other studies have not observed this reflection in V$_{1/3}$NbS$_2$.[19,33] Recognizing the existence of this $R\bar{3}c$ structure now provides an explanation for the out-of-plane periodicity if some crystals contained a substantial fraction of $R\bar{3}c$; in this work, it serves as means to further delimit possible noncollinear magnetic configurations. This propagation vector, which implies ferromagnetic order in $ab$, along with the absence of AHE and birefringence, is only consistent with a 120º spin-spiral along $c$ (Figure 4c), though we cannot distinguish between a clockwise or counter-clockwise spiral (Table S5). In agreement with low carrier densities of Figure 3c, DFT calculations for the 120º spin spiral (Figure S7) show a Fermi level lying close to the top of the hole pocket at Γ. Future studies will explore these electronic details.

In this study, we discover that V$_{1/3}$NbS$_2$ can crystallize in the conventional $P6_322$ structure or a $R\bar{3}c$ space group that has not been previously isolated. Both structures retain the in-plane $\sqrt{3} \times \sqrt{3}$ superlattice but are distinguished by their out-of-plane intercalant periodicity. Overall, magnetic properties of both $P6_322$ and $R\bar{3}c$ crystals are consistent with highly-compensated antiferromagnets with magnetic moments in the $ab$ plane. Yet, the conversion from a metallic altermagnet ($P6_322$) to a semimetallic noncollinear AFM ($R\bar{3}c$) shows that out-of-plane superlattices can be used to manipulate the magnetism in intercalation compounds. The potential for a reversible transition between these intercalant orders via external stimuli (*e.g.*, pressure, strain, electrochemistry) could even provide avenues for functional phase change devices that leverage tunability of magnetic/topological states.[16,34,35] In addition, while topological features such as Weyl points have been reported or predicted in this family of materials,[36,37] carrier densities are usually too high to support exotic transport. Here, the $R\bar{3}c$ structure displays surprisingly low carrier density, suggesting that synthesizing other $T_xMCh_2$ materials in this structure can be a powerful tool for co-designing topology and magnetism in intercalated TMD superlattices.

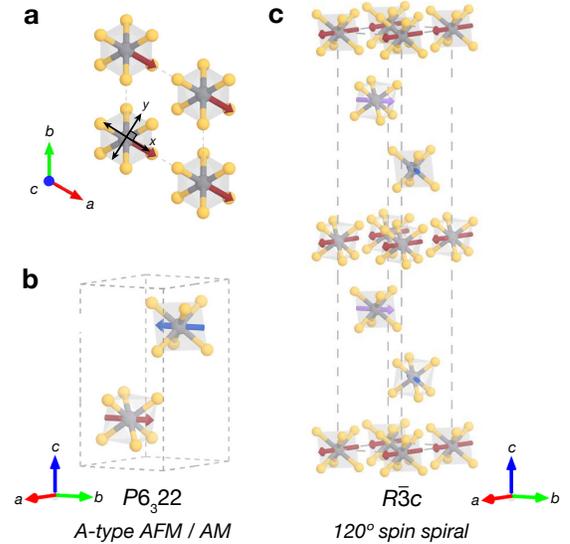

**Figure 4. a)** Coordinates for magnetic space group analysis of $P6_322$ and $R\bar{3}c$ V$_{1/3}$NbS$_2$ single crystals. **b)** A-type AFM magnetic structure for $P6_322$ V$_{1/3}$NbS$_2$. **c)** Proposed 120° spin spiral structure for $R\bar{3}c$ V$_{1/3}$NbS$_2$.

## AUTHOR INFORMATION

### Corresponding Author

D. Kwabena Bediako – Department of Chemistry, University of California, Berkeley, orcid.org/0000-0003-0064-9814, Email: bediako@berkeley.edu

### Present Addresses

† Lilia S. Xie: Department of Chemistry and Princeton Materials Institute, Princeton University, Princeton, NJ 08540, USA.

### Notes

The authors declare no competing financial interests.

## ACKNOWLEDGMENT

We are grateful to M. Frontzek for helpful discussions. We extend our thanks to J. Vigil and G. Hegel for assistance with PXRD measurements, as well as H. Jayakumar and J. Liang for assistance with XPS measurements. This material is based upon work supported by the US National Science Foundation, under award no. 2426144 (DKB). YP acknowledges support from the NSF through the University of Wisconsin Materials Research Science and Engineering Center (DMR-2309000). Experimental work at LBNL and UC Berkeley was funded by the Quantum Materials (KC2202) program under the U.S. Department of Energy, Office of Science, Office of Basic Energy Sciences, Materials Sciences and Engineering Division under Contract No. DE-AC02-05CH11231 (JO). VS and JO received support from the Gordon and Betty Moore Foundation's EPiQS Initiative through Grant GBMF4537 to JO at UC Berkeley. OG acknowledges support from the US National Science Foundation Graduate Research Fellowship Program for a predoctoral fellowship (grant no. DGE 1752814). DKB also acknowledges support from the Heising–Simons Faculty Fellowship and the Philomathia Foundation. BHG was supported by the Schmidt Science Fellows in partnership with the Rhodes Trust and the Max Planck Society. LB and DAM acknowledge support by the NSF Platform for the Accelerated Realization, Analysis, and Discovery of Interface Materials (PARADIM) under cooperative agreement No. DMR-2039380. Electron microscopy was supported by supported by the Platform for the Accelerated Realization, Analysis, and Discovery of Interface Materials (PARADIM) under NSF Cooperative Agreement no. DMR-2039380. This work made use of the Cornell Center for Materials Research (CCMR) Shared Facilities. The FEI Titan Themis 300 was acquired through No. NSF-MRI-1429155, with additional support from Cornell University, the Weill Institute, and the Kavli Institute at Cornell. The Thermo Fisher Helios G4 UX FIB was acquired with support by NSF No. DMR-1539918. The Thermo Fisher Spectra 300 X-CFEG was acquired with support from PARADIM, an NSF MIP (DMR-2039380), and Cornell University. SCXRD and XPS work at the Molecular Foundry was supported by the Office of Science, Office of Basic Energy Sciences, of the U.S. Department of Energy under Contract No. DE-AC02-05CH11231.

## ABBREVIATIONS

AFM, antiferromagnet; AHE, anomalous Hall effect; TMD, transition metal dichalcogenide; HAADF-STEM, high-angle annular dark field scanning transmission electron microscopy.